# Beam test of CSES silicon strip detector module

Da-Li Zhang (张大力)[1,2;1]    Hong Lu (卢红)[2]    Huan-Yu Wang (王焕玉)[2]    Xin-Qiao Li (李新乔)[2]    Yan-Bing Xu (徐岩冰)[2]    Zheng-Hua An (安正华)[2]    Xiao-xia Yu (于晓霞)[2]    Hui Wang (王辉)[2]    Feng Shi (石峰)[2]    Ping Wang (王平)[2]    Xiao-Yun Zhao (赵小芸)[2]

[1] University of Chinese Academy of Sciences (UCAS), Beijing 100049, China

[2] Institute of High Energy Physics (IHEP), Chinese Academy of Sciences (CAS), Beijing 100049, China

**Abstract**: The silicon-strip tracker of the China Seismo-Electromagnetic Satellite (CSES) consists of two double-sided silicon strip detectors (DSSDs) which provide incident particle tracking information. The low-noise analog ASIC VA140 was used in this study for DSSD signal readout. A beam test on the DSSD module was performed at the Beijing Test Beam Facility of the Beijing Electron Positron Collider (BEPC) using a 400~800 MeV/c proton beam. The pedestal analysis results, RMSE noise, gain correction, and intensity distribution of incident particles of the DSSD module are presented.

**PACS**: 07.87.+v, 29.40.Gx, 07.05.Hd    **DOI**:    10.1088/1674-1137/41/5/056101

## 1 Introduction

There have been some studies on the possible correlation between particle bursts and earthquakes using data gathered by high energy particle detectors on satellites such as NOAA [1], PETS-SAMPEX [2], the MIR orbital station, METEOR-3, and GAMMA [3]. The China Seismo-Electromagnetic Satellite (CSES) was proposed in 2003 [4] as the first Chinese space-based geophysical field observation satellite system, and is scheduled to be launched in 2017. The primary objectives of the CSES are to monitor and analyze seismo-ionospheric perturbation in inner Van Allen radiation belt. By using this effect, the system serves as a new approach to short-term earthquake forecast via satellite observation.

The CSES monitors seismo-ionospheric perturbation features by detecting electromagnetic field, plasma parameters, and energetic particles. The High Energy Particle Package (HEPP) is one of the eight payloads on CSES. The HEPP measures the high-energy charged particle bursts of protons and electrons that may be associated with earthquakes. It measures proton fluxes in the energy range from 2 to 200 MeV and the electron energy between 0.1 and 50 MeV with two sub-payloads: HEPP of the low energy range (HEP-L) and HEPP of the high energy range (HEPP-H). The HEPP-H is built in the form of a particle telescope and is comprised of four sub-detectors: the silicon-strip tracker (STK), plastic scintillator detector (PS detector), CsI (Tl) calorimeter, and anticoincidence plastic scintillation detector(ACD). The geometrical acceptance of HEPP-H is 75 cm2sr. Fig. 1 shows a sketch of the HEPP-H.

STK was originally developed for accelerator and collider experiments such as LHCb [5]. STK has the advantage of high hit resolution and low-noise performance in the AMS space experiment [6]. The STK of HEPP-H consists of two layers of double-sided silicon strip detectors (DSSDs) and the required typical angular resolution is 5°. It mainly provides the tracking information of the incident particles. Four layers of CsI (Tl) calorimeters measure the deposited energy of incident particles. The

*Supported by The XXX Civil space programme
1) E-mail: zhangdl@ihep.ac.cn

PS detector provides the trigger for the whole system, while the STK, PS detector, and CsI (TI) calorimeter together work as a ΔE-E telescope to achieve the typical particle identification of 90%. The ACD system consists of five plastic scintillators that surround the spectrometer. It detects and excludes particles that are outside the detective field or that traverse the spectrometer.

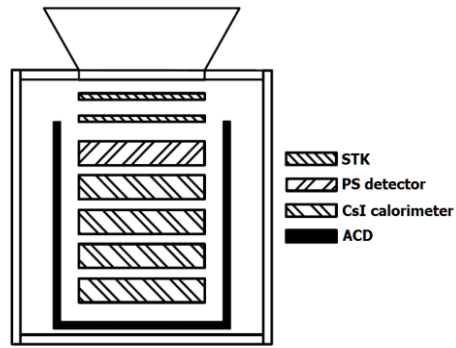

Fig. 1.　Cross-sectional view of HEPP-H.

DSSDs are commonly used as tracker in accelerator and collider experiments, such as the CMS experiments conducted at the LHC [7] and Belle II at the SuperKEKB collider [8]. In the space astrophysics detection field, DSSDs provide tracking information, such as in the AMS [6], MEGA [9], and ASTRO-H [10]. In China, the HEPP-H is the first space payload to use DSSDs as tracker.

Performance tests on the DSSDs of the AMS have been performed with 50 GeV electrons at CERN and with light ions at GSI [11]. In this study, a test was performed on the CSES DSSD module with a proton beam of 400~800 MeV/c at the Beijing Test Beam Facility (Beijing-TBF). The primary goals of beam test included the following:

　　1. To validate the prototype low-power front-end electronics used to read out the DSSD;

　　2. To investigate the pedestal, root-mean-square error (RMSE) noise, and gain factor of the DSSD module;

　　3. To draw the intensity distribution of incident particles by using the corrected data of DSSD beam test.

Section 2 briefly describes the setup of the beam test and DSSD module. The beam test results are discussed in Section 3, and conclusions are drawn in Section 4.

## 2 Experiment setup

### 2.1 Beamline

The experiment was performed at the Beijing-TBF. The Beijing-TBF is running on the linear electron accelerator (LINAC) of BEPC in parasitical mode and is a beam line of particles with middle energy [12]. The E3 line is a secondary beamline produced by an electron impinging target [13]; Be was used as the target in this experiment. The secondary particle beam mainly consists of e±, π±, and p, with a momentum of 400~800 MeV/c and a momentum resolution of 1% [14]. The repetition rate is around 10 Hz and the beam spot is around 1 cm (core) in size with a few more centimeters including the tail. The DSSD module was tested using 400~800 MeV/c protons. A schematic diagram of the beamline layout is shown in Fig. 2.

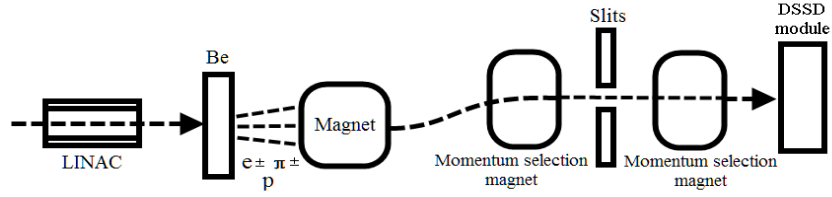

Fig. 2　Schematic layout of the E3 beamline.

## 2.2 DSSD module and DAQ

The double-sided silicon strip detectors of the CSES were manufactured by Micronsemiconductor (type TTT2 (DS)-300). They are composed of 295 μm thick n-type material with an active area of 100 mm×100 mm. Either side of the DSSD (ohmic-side and junction-side) has 128 active micro-strips of 760 μm pitch and 700 μm width. To reduce the number of the read-out channels of the DSSD, every four strips are merged into one channel during application. Once the DSSD module was triggered, only the highest pulse height signal of each detective side was recorded. For HEPP-H, the spatial resolution of the 1-strip cluster position distribution can be defined by $p/\sqrt{12}$ [15], where p is the width of each channel (3.04 mm) and the spatial resolution is 0.888mm.

The depletion voltage of the DC-coupled TTT2 is -30V. The high-voltage source (HV) provides -40V for the DSSD to reach full depletion. Fig. 3 shows where RC filters are applied for the HV and each channel of DSSD is coupled with the front-end electronics by a resistor and capacitor. To simplify the DSSD electronics readouts, an ASIC (VA140) is applied to amplify, shape, and hold the DSSD signals. The VA140 has 64 channels which amplify the DSSD charge pulse. The charge mode of the VA140 must be set according to the output charge of the ohmic-side and junction-side of the DSSD. The VA140 readout is converted to a digital signal by the ADC for offline data analysis.

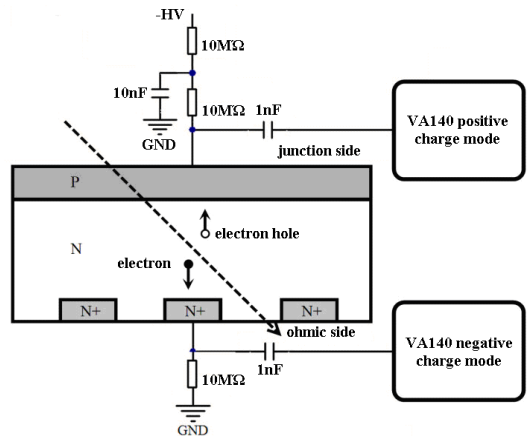

Fig. 3.　Coupling circuit of one DSSD channel.

A schematic diagram of the DSSD module beam test and DAQ is shown in Fig. 4. The DSSD and read-out electronics are placed in an aluminum box to shield them from light and electromagnetic interference. The sensor surface of the DSSD is perpendicular to the incident beam direction; a plastic scintillator situated behind the DSSD functions as the event trigger. When the beam particles traverse the DSSD and hit the plastic scintillator (PS), the output signal of the PS is discriminated and shaped into a TTL pulse. As soon as this TTL pulse triggers the FPGA, the FPGA controls the VA140 to shape and hold the DSSD signals. The VA140 outputs are digitalized by the AD9243, then the data is transmitted to the PC through the RS232 serial port. To evaluate the VA140 performance, a radioactive source test [16] on the silicon detectors was performed prior to the beam test.

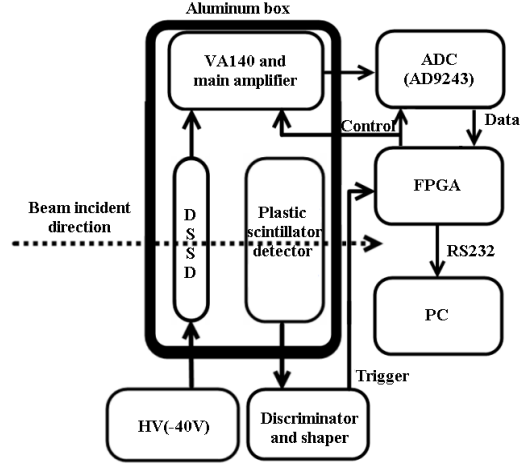

Fig. 4. Schematic of the DSSD module beam test and DAQ.

The DSSD module was tested using a 400-800 MeV/c proton beam. The proton beam covers a kinetic energy range of 81.7-295 MeV. (The HEPP covers a proton energy range of 15-200 MeV.) The beam test, by design, covers a large portion of the HEPP-H's proton detection range.

## 3. Results

### 3.1 Pedestal analysis

The data acquisition system of the DSSD module was run 38,070 times to evaluate the pedestal and RMSE noise of each channel when the particle beam was off. The mean value of each channel represents the pedestal, while the RMSE noise of each strip was computed on an event-by-event basis as follows:

$$\sigma = \sqrt{\frac{1}{N}\sum_{i=1}^{N}(Raw_i - P_i)^2} \cdot \quad (1)$$

where i is the event number, N is the total number of events in the full run time, Rawi represents the raw data, and the pedestal (Pi) was computed [17].

Fig. 5 shows the pedestals and RMSE noise of the 64 DSSD channels. The junction-side channels correspond to channels 1-32 and the ohmic-side channels correspond to channels 33-64. Noisy and dead channels were observed, as shown in Fig. 5(b); the relative high RMSE channels (5, 18, 51, 52) were noisy, and the extremely low RMSE noise channels (channels 15, 32, 64) were not connected to the front-end electronics (i.e., were dead) because electronic system noise (around 1 ADC channel) was much lower than intrinsic detector noise. SThe RMSE noises of junction-side were 4-5 ADC channels and ohmic-side were 9-11 except the noisy channels. The RMSE noises differences between the junction-side and ohmic-side come from the different coupling circuits and ASIC gain factors of each detective side. By using the gain factors of section 3.2, the RMSE noises of junction-side were 18.6-25.8 keV and ohmic-side were 20.1-31.1 keV (the minimum deposited energy in DSSD of HEPP-H design is 50 keV). The RMSE noise will be greatly reduced by cleaning the surface of DSSD and adding shield when the full HEPP-H was assembled. These results were used for two-dimensional intensity distribution drawing to substrate the pedestal and cut noise.

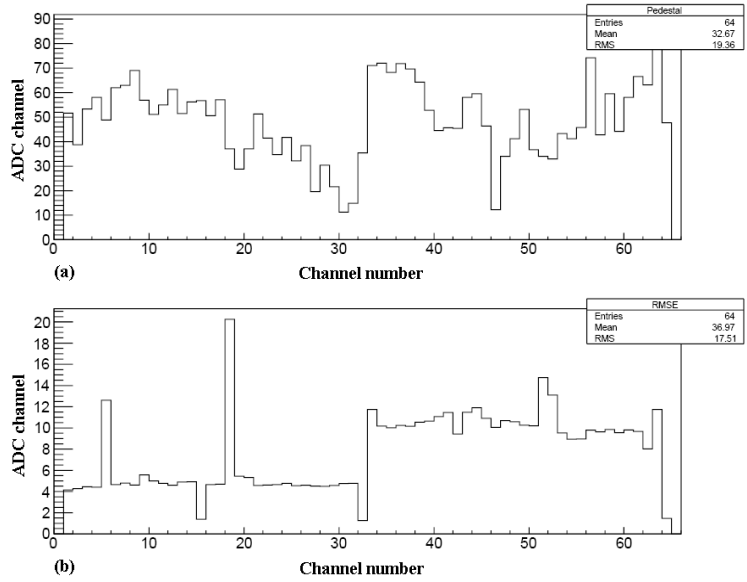

Fig. 5. (a) Pedestal profile histogram. (b) RMSE noise distributions

### 3.2 Electronic gain factor analysis

The data of each channel was corrected to account for the different gain factors among channels. The DSSD module was also tested using the proton beam to evaluate the gain factors. Fig. 6 shows the pulse height histograms of channel 16 (junction-side) at five beam momentums: 400, 500, 600, 700, and 800 MeV/c. As momentum increased, the incident particle deposited much less energy linked to the energy loss around the minimum ionization [18]. Taking into account the measuring error of the readout system, the experimental energy loss distributions in Fig. 6 were fitted by the Landau probability density convolved with a Gaussian [19] and the most probability values (MPV) of the charge deposition were determined accordingly.

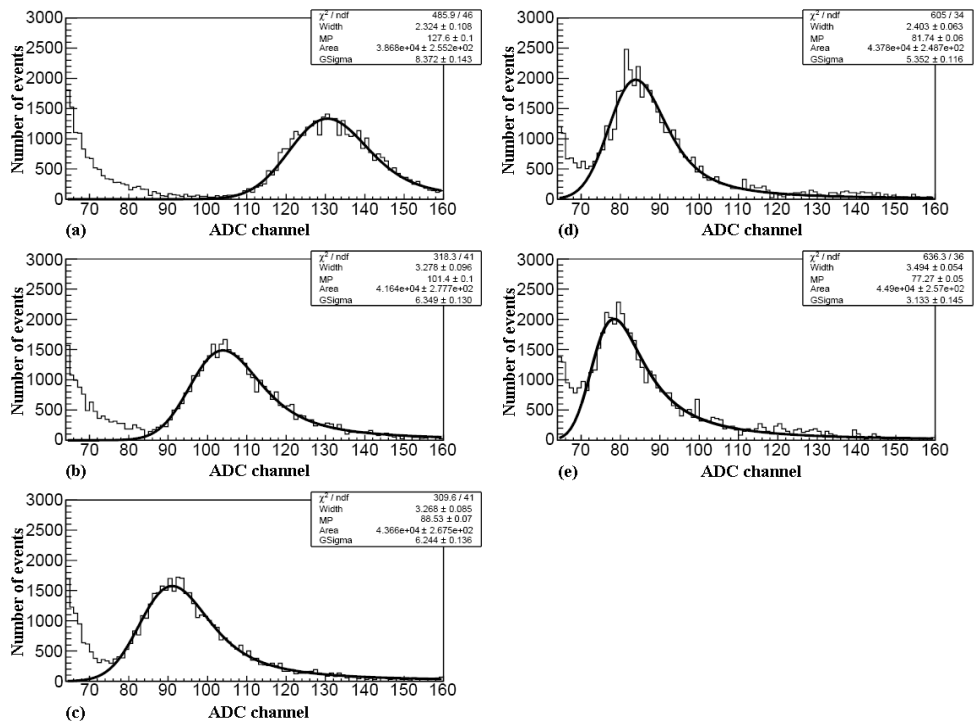

Fig. 6. Pulse height histograms of channel 16 at five beam momenta: (a) 400 MeV/c (MPV=128). (b) 500 MeV/c (MPV=101). (c) 600 MeV/c(MPV=88.5) (d) 700 MeV/c(MPV=81.7). (e) 800 MeV/c (MPV=77.3).

The electronic gain factors were determined by analyzing the measured MPV and the simulated MPV. Simulated MPV derived from Monte Carlo simulation (Geant4 Simulation Toolkit) of a proton beam normally incident on a 300-μm DSSD strip. In Fig. 7, the x coordinate is the MPV of the experimental energy loss distributions and the y coordinate is the MPV of the simulated energy loss. The electronic gain factor was determined as the p0 (slope of the liner fitting) in Fig. 7. The electronic gain factors of other channels were obtained similarly. The gain factors of the junction-side were 4.40-5.24 keV per ADC channel and those of the ohmic-side were 2.20-2.94 keV per ADC channel (excepting the dead channels.)

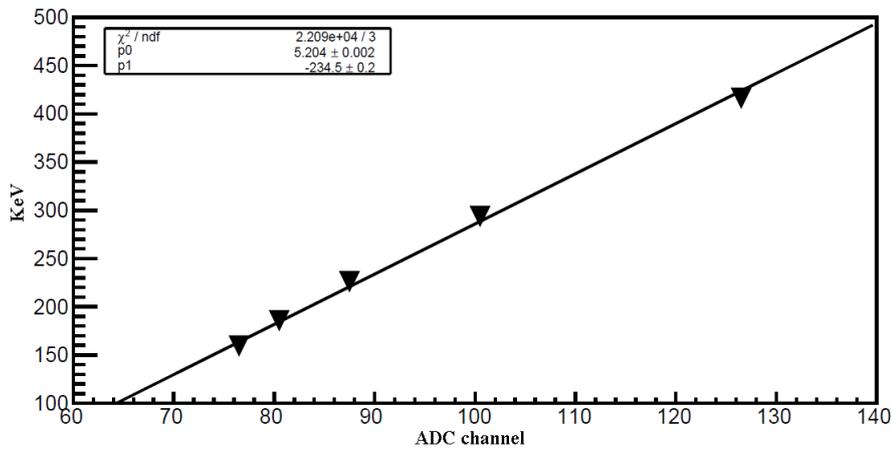

Fig.7 Measured MPV as described in Fig.6, compared with MPV derived from Monte Carlo simulation.

**3.3 Pedestal subtraction and gain correction of data**

To calculate the normalized pulse height histograms, the ADC data of each channel was computed on a channel by channel basis as

$$D_i = (Raw_i - P_i) \times G_i / G_{min} . \qquad (2)$$

where i represents the channel number, $Raw_i$ represents the raw data, $P_i$ is the pedestal, the minimum gain factor ($G_{min}$) was identified among the 32 channels of ohmic or junction side, and the gain factor ($G_i$) was computed from the data under the beam (Section 3.2). Fig. 8 shows the corrected pulse height histograms of 64 channels under the proton beam of 500 MeV/c after pedestal subtraction and gain correction. The strips near the beam spot center had higher counts; channel 16 (panel 16) is a good example. Channel 15, 32, 64 were dead and channel 18, 51 were cut by threshold. The corrected data were used in the subsequent drawing of incident particle intensity distribution.

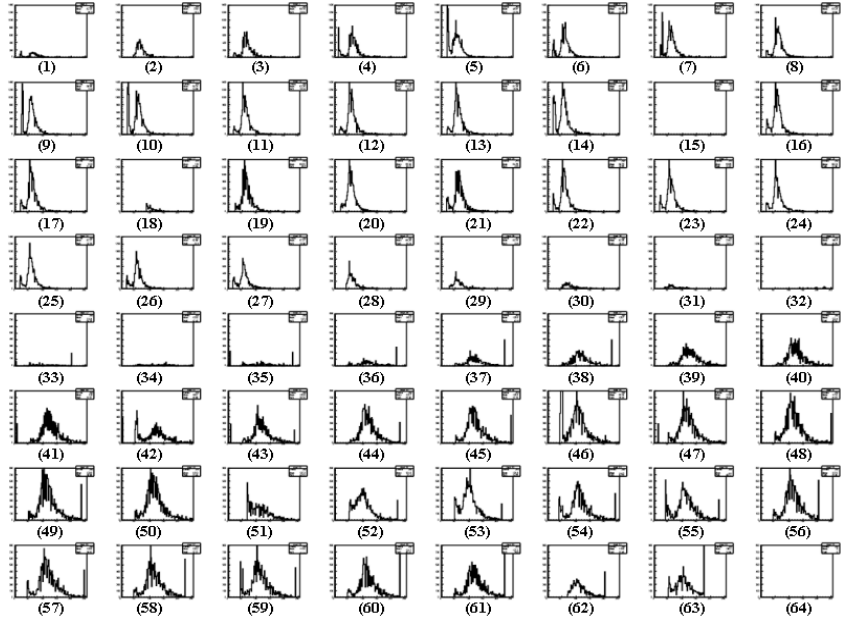

Fig. 8. Pulse height histogram of junction side (panels 1-32) and ohmic side (panels 33-64) at 500MeV/C proton.

**3.4 Intensity distribution of incident particles**

Particle incident positions were drawn according to the beam test data corrected via pedestal subtraction and gain correction (Eq. (2)). Event selection was performed according to the RMSE noise data (Fig. 5(b)). An event was considered valid when the pulse height of the hit strip exceeded $4\sigma$. A single particle event was recorded only when exactly one ohmic-side strip and one junction-side strip were hit. The ohmic-side strips were oriented to measure the x coordinates and the junction-side strips were oriented to measure the y coordinates; the corresponding particle incident position was reconstructed from the X-Y coordinates of hit strips. Fig. 9 shows the intensity distribution of incident particles in a two-dimensional color contour format using proton beams of 400, 500, 600, 700, and 800 MeV/c. The beam repetition rate is highest at 600 MeV/c with the highest number of events (Fig. 9(c)).

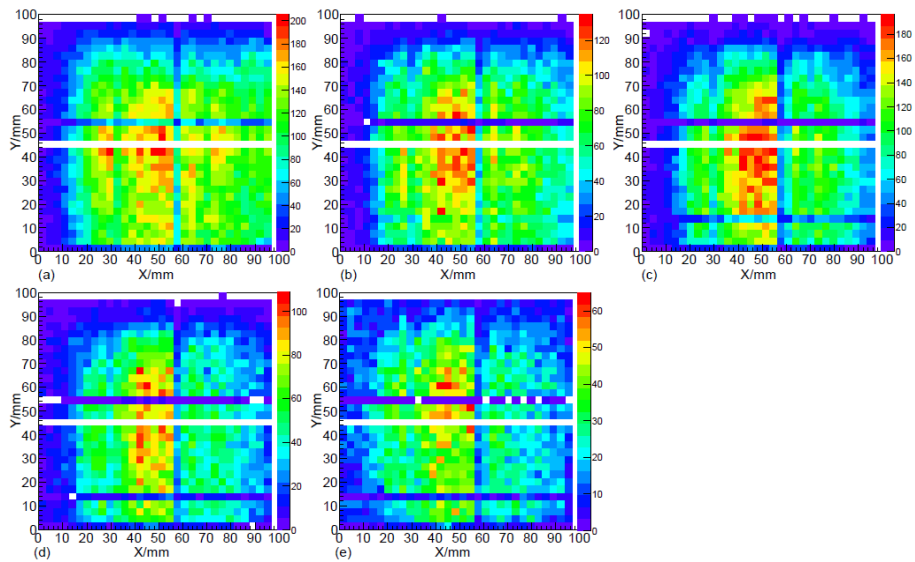

Fig. 9. Intensity distribution of incident particles: (a) 400 MeV/c proton (b) 500 MeV/c proton (c)

600 MeV/c proton (d) 700 MeV/c proton (e) 800 MeV/c proton.

As momentum increased, the incident particle deposited less energy in the DSSD. Fig.10 shows the beam profile as measured in the y and x coordinate by the DSSD module and the previous beam distributions measured by the MWPC of this beamline were introduced in Ref. [14]. Compared with the results of Fig. 5(b), the counts of the noisy channels (channel 5, 18, 51, 52) were cut by threshold. The dead channels (channel 15, 32, 64) can also be observed in Fig.10. Although some strips were damaged, the beam spot pattern was drawn successfully after data correction and event selection.

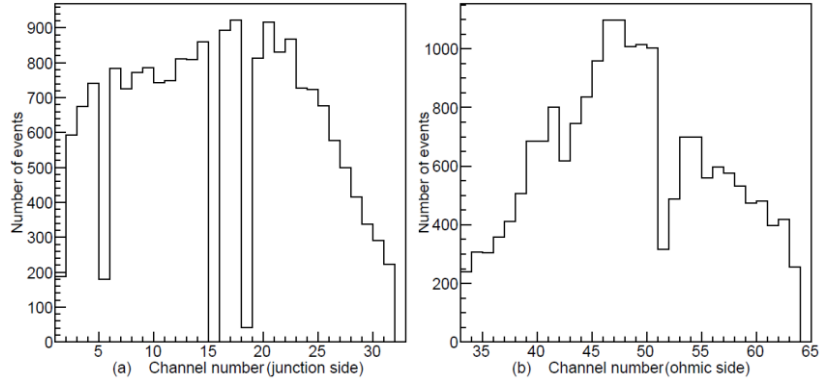

Fig.10. Measured beam profile of 800 MeV/c proton: (a) Profile measured in the y coordinate (b) Profile measured in the x coordinate. The width of each channel is 3.04 mm.

## 4 Conclusion

The CSES silicon-strip tracker provides vital tracking information for the HEPP-H. In this study, a beam test on the DSSD module was performed using a proton beam of 400~800 MeV/c at the Beijing-TBF. The design of the readout electronics of DSSD was successfully validated by the test results. The pedestal and RMSE properties of the DSSD module were also investigated; the RMSE noises of all channels are 18.6 to 31.1 keV. The electronic gain factors of all channels were determined to be 2.20-5.24 keV per ADC channel. The intensity distributions of incident proton were drawn in a two-dimensional color contour format by using the corrected data. These results provide useful information for following DSSD tests performed at this beamline.

A follow-up beam test on a full HEPP-H is planned, during which time the authors will investigate the energy calibration and angular resolution of the STK.

*Acknowledgements:*

We would like to express our appreciation to the staff of the Beijing Test Beam Facility for their valuable assistance throughout the beam test process.

**References**

1  Cristiano Fidani et al, Remote Sensing, **2**: 2170-2184 (2010)

2  V. Sgrigna et al, Journal of Atmospheric and Solar-Terrestrial Physics, **67**:1448-1462 (2005)

3  S. Yu. Aleksandrin et al, **21**: 597–602 (2003)

4  Shen Xuhui, Chinese Journal of Space Science, **34**(5): 558-562 (2014)

5  Mark Tobin et al, Nuclear Instruments and Methods in Physics Research A, **831**: 174-180 (2016)

6  Maurice Bourquin et al, Nuclear Instruments and Methods in Physics Research A, **541**: 110–116 (2005)

7  S. MY et al, Nuclear Instruments and Methods in Physics Research A, **446**: 229–234 (2000)

8  K. Adamczyk et al, Nuclear Instruments and Methods in Physics Research A, **831**: 80–84 (2016)

9  G. Kanbach et al, Nuclear Instruments and Methods in Physics Research A, **541**: 310–322 (2005)

10  M. Kokubun et al, Nuclear Instruments and Methods in Physics Research A, **623**: 425–427 (2010)

11  G. Ambrosi et al, Nuclear Instruments and Methods in Physics Research A, **435**: 215–223 (1999)

12  LI Jia-Cai et al. HEP & NP, **28**(12): 1269—1277 (2004) (in Chinese)

13  http://english.ihep.cas.cn/rh/rd/dep/sywlbtbf/sywlbtbf_articles/201110/W020120606406778866830.pdf, retrieved 10th November 2010

14  http://english.ihep.cas.cn/rh/rd/dep/sywlbtbf/sywlbtbf_articles/201110/W020111012419711925365.pdf, retrieved 10th November 2010

15  T.W Versloot et al, LHCb 2007-119 (2007)

16  ZHANG Da-Li et.al, Nuclear Electronics & Detection Technology, **35**(11):1150-1153 (2015) (in Chinese)

17  A. Berra et al, Nuclear Instruments and Methods in Physics Research A, **798**: 80–87 (2015)

18  J. A. Sauvaud et al, Planetary and Space Science, **54**: 502-511 (2006)

19  S. Hancock et al, Phys. Rev. A, **28**: 615–620 (1983)